\documentclass[cits]{PoS}

\newcommand{\RTau}[1]{R_{\tau, \mbox{\tiny #1}}}
\newcommand{\ind}[2]{^{\mbox{\scriptsize $#1$}}_{\mbox{\scriptsize #2}}}
\newcommand{\inds}[2]{^{\mbox{\tiny $#1$}}_{\mbox{\tiny #2}}}
\newcommand{\Nc}{N_{\mbox{\scriptsize c}}}
\newcommand{\Vud}{V_{\mbox{\scriptsize ud}}}
\newcommand{\Sew}{S_{\!\mbox{\tiny EW}}}
\newcommand{\dpew}{\delta'_{\mbox{\tiny EW}}}
\newcommand{\DeltaQCD}[1]{\Delta^{\mbox{\tiny #1}}_{\mbox{\tiny QCD}}}
\newcommand{\DeltaHad}[1]{\delta^{\mbox{\tiny #1}}_{\mbox{\scriptsize had}}}
\newcommand{\va}{_{\mbox{\tiny V/A}}}
\newcommand{\ML}{M_{l}}
\newcommand{\MTau}{M_{\tau}}
\newcommand{\nf}{n_{\mbox{\scriptsize f}}}
\newcommand{\zva}{\zeta\va}
\newcommand{\DP}{\Delta\Pi}
\newcommand{\includeplots}[2]{%
   \centerline{\includegraphics[width=62.5mm]{#1}%
   \hspace{15mm}%
   \includegraphics[width=62.5mm]{#2}}}

\title{Inclusive tau~lepton decay: \\ the effects due to hadronization}
\ShortTitle{Inclusive tau~lepton decay: the effects due to hadronization}

\author{\speaker{A.V.~Nesterenko}\\
        Bogoliubov Laboratory of Theoretical Physics,
        Joint Institute for Nuclear Research,\\
        Joliot Curie 6, Dubna, Moscow region, 141980, Russian Federation\\
        E-mail: \email{nesterav@theor.jinr.ru}}

\abstract{The inclusive $\tau$~lepton hadronic decay and its description
within Dispersive approach to Quantum Chromodynamics are briefly discussed.}

\FullConference{Xth Quark Confinement and the Hadron Spectrum\\
                 8-12 October 2012\\
                 TUM Campus Garching, Munich, Germany}

\begin{document}

\section{Introduction}

The process of the inclusive $\tau$~lepton decay into hadrons constitutes
a unique opportunity to explore the nonperturbative nature of the strong
interaction at low energies. The experimental data on hadronic
$\tau$~decay are commonly employed in various tests of Quantum
Chromodynamics~(QCD) and entire Standard Model, that puts strong
limits on possible New Physics beyond the latter.

The pertinent experimentally measurable quantity is the ratio of the total
width of $\tau$~lepton decay into hadrons to the width of its leptonic
decay. Usually, this ratio is decomposed into several parts, specifically
\begin{equation}
\label{RTauExp}
R_{\tau} = \frac{\Gamma(\tau^{-} \to \mbox{hadrons}^{-}\,
\nu_{\tau})} {\Gamma(\tau^{-} \to e^{-}\, \bar{\nu}_{e}\, \nu_{\tau})} =
\RTau{V}^{\mbox{\tiny $J$=0}} + \RTau{V}^{\mbox{\tiny $J$=1}} +
\RTau{A}^{\mbox{\tiny $J$=0}} + \RTau{A}^{\mbox{\tiny $J$=1}} + \RTau{S}.
\end{equation}
In the right hand side of this equation the last term accounts for the
$\tau$~lepton decay modes which involve strange quark, whereas the other
terms account for the hadronic decay modes involving light quarks~(u, d)
only and associated with vector~(V) and axial--vector~(A) quark currents,
respectively. The superscript~$J$ indicates the angular momentum in the
hadronic rest frame.

\begin{figure}[t]
\includeplots{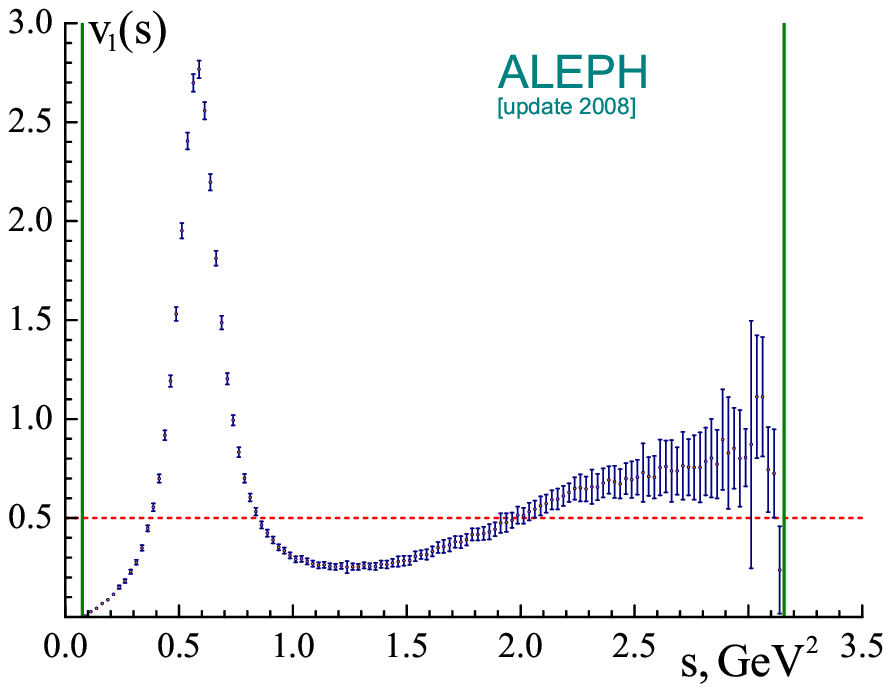}{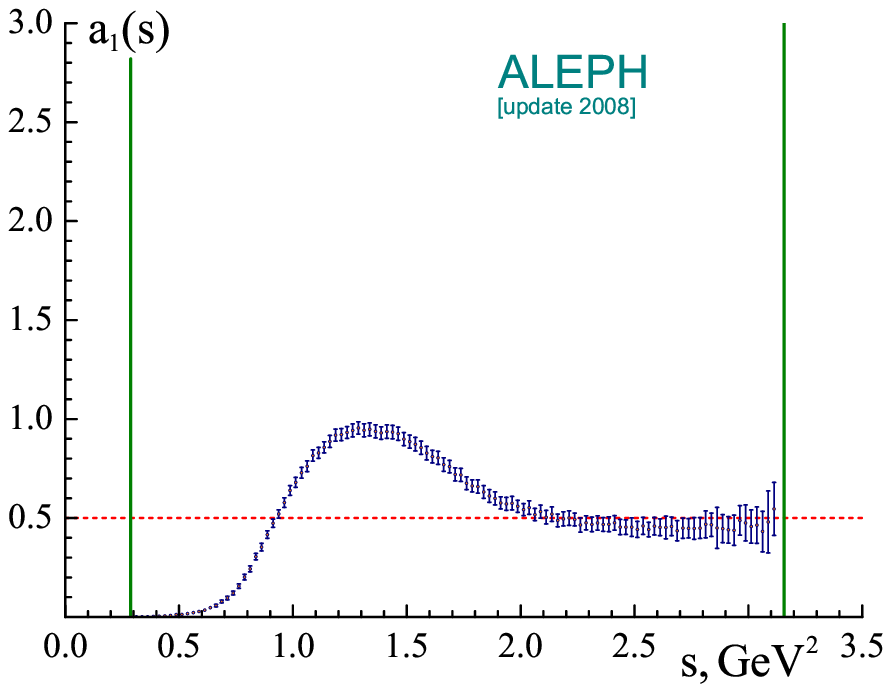}
\caption{The inclusive $\tau$~lepton hadronic decay vector (left plot) and
axial--vector (right plot) spectral functions~\cite{ALEPH9805, ALEPH0608}.
Vertical solid lines mark the boundaries of respective kinematic
intervals, whereas horizontal dashed lines denote the naive massless
parton model predictions.}
\label{Plot:ALEPHSpFun}
\end{figure}

The quantities appearing in Eq.~(\ref{RTauExp}) can be evaluated by making
use of the so--called spectral functions, which are extracted from the
experiment. For the zero angular momentum ($J=0$) the vector spectral
function vanishes (that leads to~$\RTau{V}^{\mbox{\tiny $J$=0}}=0$),
whereas the axial--vector one is commonly approximated by Dirac
$\delta$--function, since the dominant contribution is due to the pion
pole here. The experimental predictions~\cite{ALEPH9805, ALEPH0608} for
the nonstrange spectral functions corresponding to $J=1$ are presented in
Fig.~\ref{Plot:ALEPHSpFun}. In what follows we shall restrict ourselves to
the consideration of terms~$\RTau{V}^{\mbox{\tiny $J$=1}}$
and~$\RTau{A}^{\mbox{\tiny $J$=1}}$ of $R_{\tau}$--ratio~(\ref{RTauExp}).

The aforementioned quantities can be represented in the following form
\begin{equation}
\label{RTauTheor}
\RTau{V/A}^{\mbox{\tiny $J$=1}} = \frac{\Nc}{2}\,|\Vud|^2\,\Sew\,
\Bigl(\DeltaQCD{V/A} + \dpew \Bigr) ,
\end{equation}
where $\Nc=3$ is the number of colors, $|\Vud| = 0.97425 \pm 0.00022$ is
Cabibbo--Kobayashi--Maskawa matrix element~\cite{PDG2012}, $\Sew = 1.0194
\pm 0.0050$ and $\dpew = 0.0010$ stand for the electroweak corrections
(see Refs.~\cite{BNP, EWF}), and
\begin{equation}
\label{DeltaQCDDef}
\DeltaQCD{V/A} = 2\int_{m\va^2}^{\ML^2}\!\!
\biggl(1-\frac{s}{\ML^2}\biggr)^{\!\!\!\!2}\!\biggl(1+2\frac{s}{\ML^2}\biggr)
R^{\mbox{\tiny V/A}}(s)\,
\frac{d s}{\ML^2}
\end{equation}
denotes the QCD contribution to Eq.~(\ref{RTauTheor}). In the integrand of
Eq.~(\ref{DeltaQCDDef})
\begin{equation}
\label{RDef}
R(s) = \frac{1}{2 \pi i} \lim_{\varepsilon \to 0_{+}}
\Bigl[\Pi(s+i\varepsilon)-\Pi(s-i\varepsilon) \Bigr],
\end{equation}
where $\Pi(q^2)$ is the hadronic vacuum polarization function
\begin{equation}
\label{PDef}
\Pi_{\mu\nu}(q^2) = i\!\int\!d^4x\,\,e^{i q x} \langle 0 |\,
T\!\left\{J_{\mu}(x)\, J_{\nu}(0)\right\} | 0 \rangle =
\frac{i}{12\pi^2} (q_{\mu}q_{\nu} - g_{\mu\nu}q^2) \Pi(q^2)
\end{equation}
with $J_{\mu}(x)$ being the electromagnetic quark current (the indices
``V'' and~``A'' will only be shown when relevant hereinafter). It is
worthwhile to mention that for practical purposes it is also convenient
to deal with the so--called Adler function~\cite{Adler}
\begin{equation}
\label{AdlerDef}
D(Q^2) = - \frac{d\, \Pi(-Q^2)}{d \ln Q^2}, \qquad Q^2 =-q^2=-s.
\end{equation}

It is necessary to outline that in Eq.~(\ref{DeltaQCDDef}) $\ML$~denotes
the mass of the lepton on hand, whereas $m$~stands for the hadronic
threshold mass (i.e., the total mass of the lightest allowed hadronic
decay mode of this~lepton in the corresponding channel). The nonvanishing
value of~$m$ explicitly expresses the physical fact that $\tau$~lepton is
the only lepton which is heavy enough (\mbox{$\MTau \!\simeq\!\!
1.777\,$GeV$\,$\cite{PDG2012}}) to decay into hadrons. Indeed, in the
massless limit ($m=0$) the theoretical prediction for the QCD
contribution~(\ref{DeltaQCDDef}) to Eq.~(\ref{RTauTheor}) is nonvanishing
for either lepton (\mbox{$l = e, \mu, \tau$}). Specifically, the
leading--order term of Eq.~(\ref{DeltaQCDPert})
\mbox{$\Delta\ind{(0)}{pert}=1$} (which corresponds to the naive massless
parton model prediction for the Adler function~(\ref{AdlerPert})
$D\ind{(0)}{pert}(Q^2)=1$) does not depend on~$\ML$, and, therefore, is
the same for either lepton. In~the realistic case (i.e., when the total
mass of the lightest allowed hadronic decay mode exceeds the masses of
electron and muon, $M_{e}<M_{\mu}<m<\MTau$) Eq.~(\ref{DeltaQCDDef})
acquires non--zero value for the case of the $\tau$~lepton only.

\section{Inclusive $\tau$~lepton hadronic decay within perturbative approach}
\label{Sect:Pert}

In this Section we shall deal with the massless limit, that implies that
the masses of all final state particles are neglected~($m=0$). By making
use of definitions~(\ref{RDef}) and~(\ref{AdlerDef}), integrating by
parts, and additionally employing Cauchy theorem, the
quantity~$\DeltaQCD{}$~(\ref{DeltaQCDDef}) can be represented as
\begin{equation}
\label{DeltaQCDCauchy}
\DeltaQCD{} = \frac{1}{2\pi}\! \int_{-\pi}^{\pi}\!
D\bigl(M_{\tau}^{2}\,e^{i\theta}\bigr)
\bigl(1 + 2e^{i\theta} - 2e^{i3\theta} - e^{i4\theta}\bigr) d \theta,
\end{equation}
see, e.g., Refs.~\cite{P91DP92, BNP}. It is worth noting here that
Eq.~(\ref{DeltaQCDCauchy}) is only valid for the massless limit of
``genuine physical'' Adler function~$D\ind{}{phys}(Q^2)$, which possesses
the correct analytic properties in the kinematic variable~$Q^2$
(otherwise Eq.~(\ref{DeltaQCDCauchy}) can not be derived from
Eq.~(\ref{DeltaQCDDef})). However, in Eq.~(\ref{DeltaQCDCauchy}) one
usually directly employs the perturbative approximation for the Adler
function
\begin{equation}
\label{AdlerPert}
D(Q^2) \simeq D\ind{(\ell)}{pert}(Q^2) = 1 + \sum\nolimits_{j=1}^{\ell}
d_{j} \Bigl[\alpha\ind{(\ell)}{pert}(Q^2)\Bigr]^{j}, \qquad Q^2\to\infty,
\end{equation}
which has unphysical singularities in~$Q^2$. In this equation at the
one--loop level (i.e., for \mbox{$\ell=1$}) the strong running coupling
reads $\alpha\ind{(1)}{pert}(Q^2) = 4\pi/[\beta_{0}\,\ln(Q^2/\Lambda^2)]$,
where $\beta_{0}=11-2\nf/3$, $\Lambda$~denotes the QCD scale parameter,
$\nf$~is the number of active flavors, and~$d_{1}=1/\pi$, see
Ref.~\cite{AdlerPert4L} for the details. In what follows the one--loop
level with $\nf=3$ active flavors will be assumed. Eventually,
Eq.~(\ref{DeltaQCDCauchy}) corresponding to the perturbative Adler
function~(\ref{AdlerPert}) takes the form
\begin{equation}
\label{DeltaQCDPert}
\Delta\ind{}{pert} = \Delta\ind{(0)}{pert} +
\frac{4}{\beta_{0}}\!\int_{0}^{\pi}
\frac{\lambda A_{1}(\theta)+\theta A_{2}(\theta)}{\pi(\lambda^2+\theta^2)}
\,d\theta,
\end{equation}
where $\Delta\ind{(0)}{pert} \!=\! 1$,
$A_{1}(\theta) \!=\! 1 \!+\! 2\cos(\theta) \!-\! 2\cos(3\theta) \!-\! \cos(4\theta)$,
$A_{2}(\theta) \!=\! 2\sin(\theta) \!-\! 2\sin(3\theta) \!-\! \sin(4\theta)$,
and $\lambda = \ln \bigl( \MTau^2/\Lambda^2 \bigr)$.

\begin{figure}[t]
\includeplots{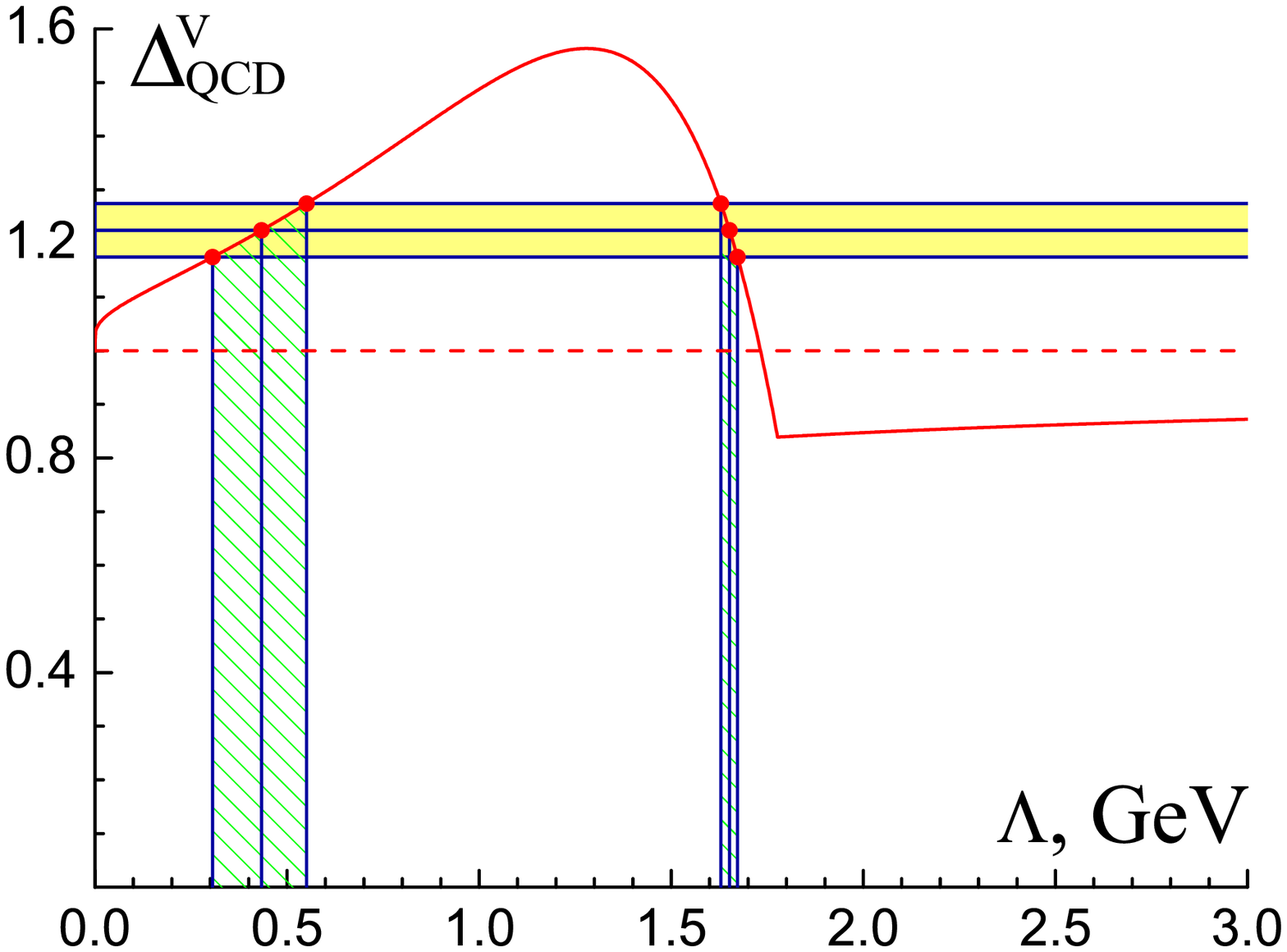}{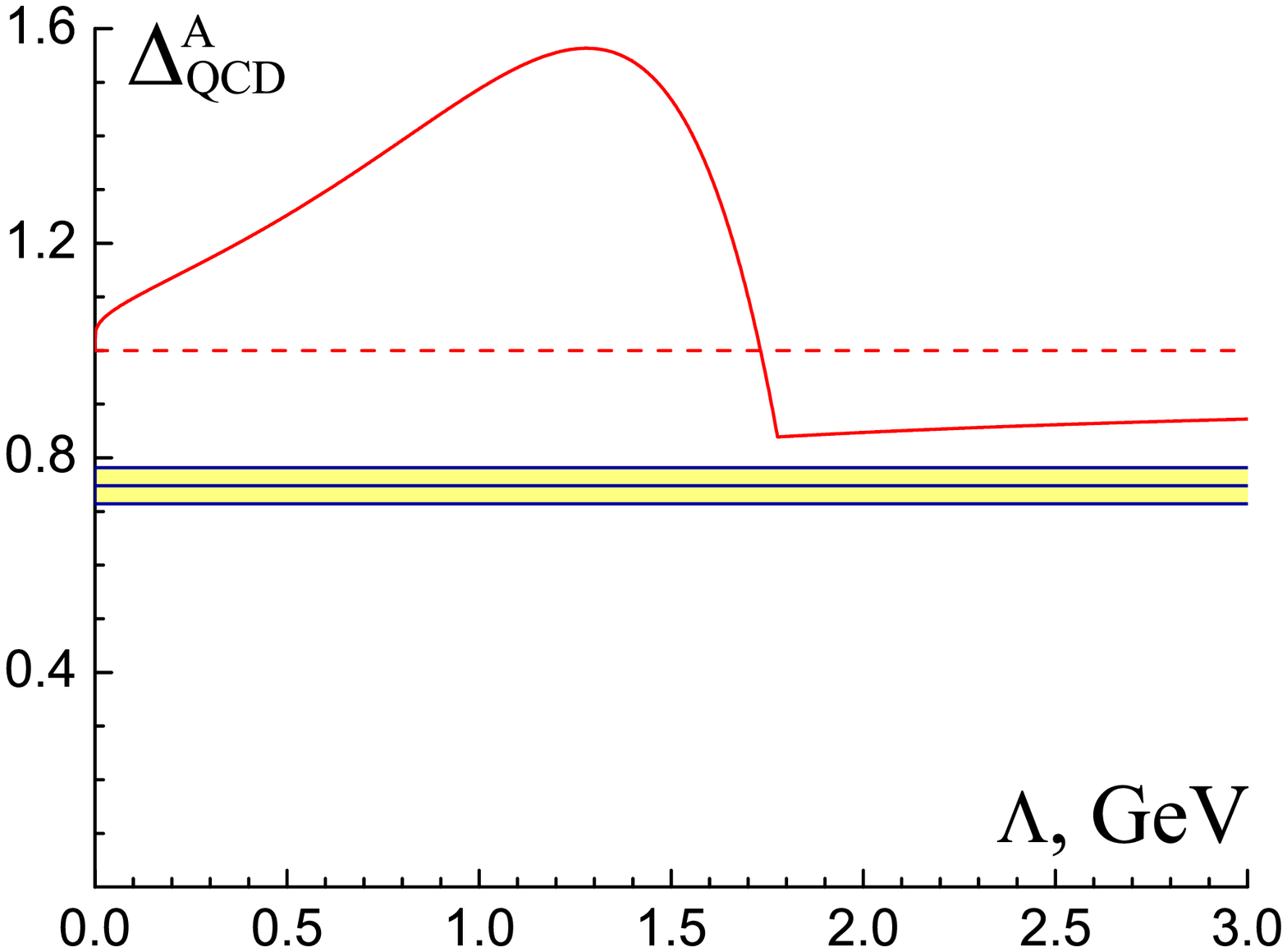}
\caption{Juxtaposition of the one--loop perturbative expression
$\Delta\inds{}{pert}$~(\protect\ref{DeltaQCDPert}) (solid curve) with
relevant experimental data~(\protect\ref{DeltaQCDExp}) (horizontal shaded
bands). The leading--order term
$\Delta\inds{(0)}{pert}=1$~(\protect\ref{DeltaQCDPert}) is denoted by
dashed line. The solution for QCD scale parameter~$\Lambda$ (if exists) is
shown by vertical dashed band.}
\label{Plot:Pert}
\end{figure}

It is worthwhile to underscore that perturbative approach provides
identical expressions~(\ref{DeltaQCDPert}) for the
functions~(\ref{DeltaQCDDef}) in vector and axial--vector channels (i.e.,
\mbox{$\Delta\ind{\mbox{\tiny V}}{pert} \equiv \Delta\ind{\mbox{\tiny
A}}{pert}$}). However, their experimental values~\cite{ALEPH9805,
ALEPH0608} are different, namely
\begin{equation}
\label{DeltaQCDExp}
\Delta\ind{\mbox{\tiny V}}{exp} = 1.224 \pm 0.050, \qquad
\Delta\ind{\mbox{\tiny A}}{exp} = 0.748 \pm 0.034.
\end{equation}
The comparison of these quantities with perturbative
result~(\ref{DeltaQCDPert}) is presented in Fig.~\ref{Plot:Pert}. As one
can infer from this figure, for the vector channel there are two solutions for
the QCD scale parameter: $\Lambda=\bigl(434^{+117}_{-127}\bigr)\,$MeV
(which is usually retained) and
$\Lambda=\bigl(1652^{+21}_{-23}\bigr)\,$MeV (which is commonly merely
disregarded). As~for the axial--vector channel, the perturbative approach
fails to describe the experimental data~\cite{ALEPH9805, ALEPH0608}, since
for any value of~$\Lambda$ the function
$\Delta\ind{}{pert}$~(\ref{DeltaQCDPert}) exceeds $\Delta\ind{\mbox{\tiny
A}}{exp}$~(\ref{DeltaQCDExp}).

\section{Dispersive approach to Quantum Chromodynamics}

It is crucial to emphasize that the presented in Section~\ref{Sect:Pert}
massless limit completely leaves out the effects due to hadronization,
which play significant role in the studies of the strong interaction
processes at low energies. Specifically, the mathematical realization of
the physical fact, that in a strong interaction process no final state
hadrons can be produced at energies below the hadronic threshold mass~$m$,
consists in the fact that the beginning of cut of corresponding hadronic
vacuum polarization function~$\Pi(q^2)$~(\ref{PDef}) in the complex
\mbox{$q^2$--plane} is located at the threshold of hadronic
production~$q^2=m^2$, but not at~$q^2=0$ (see also discussion of this
issue in Ref.~\cite{Feynman}). Such restrictions are inherently embodied
within relevant dispersion relations, which, in turn, impose stringent
physical intrinsically nonperturbative constraints on the quantities on
hand.

The complete set of dispersion relations, which express the
functions~(\ref{RDef}), (\ref{PDef}), and~(\ref{AdlerDef})
in terms of each other, reads
\begin{eqnarray}
\label{DispRelP}
\DP(q^2,\, q_0^2) &=& (q^2 - q_0^2) \int_{m^2}^{\infty}
\frac{R(\sigma)}{(\sigma-q^2)(\sigma-q_0^2)}\, d\sigma =
- \int_{-q_0^2}^{-q^2} D(\zeta) \frac{d\zeta}{\zeta}, \\
\label{DispRelD}
D(Q^2) &=& - \frac{d\,\Pi(-Q^2)}{d\,\ln Q^2} =
Q^2 \int_{m^2}^{\infty} \frac{R(\sigma)}{(\sigma+Q^2)^2}\, d\sigma, \\
\label{DispRelR}
R(s) &=& \frac{1}{2 \pi i} \lim_{\varepsilon \to 0_{+}}
\Bigl[\Pi(s+i\varepsilon)-\Pi(s-i\varepsilon) \Bigr] =
\frac{1}{2 \pi i} \lim_{\varepsilon \to 0_{+}}
\int_{s+i\varepsilon}^{s-i\varepsilon} D(-\zeta)
\frac{d\zeta}{\zeta}, \qquad \quad
\end{eqnarray}
where $\DP(q^2,\, q_0^2) = \Pi(q^2) - \Pi(q_0^2)$ and $s=q^2=-Q^2$
(see Refs.~\cite{Adler, RKP82}). For
practical purposes, it proves to be convenient to deal with the integral
representations, which express the aforementioned functions in terms of
the common spectral density~$\rho(\sigma)$. Such representations have been
derived in the framework of Dispersive approach to QCD (see
Refs.~\cite{DQCD1, DQCD2} for the details):
\begin{eqnarray}
\label{P_DQCD}
\DP^{(\ell)}(q^2,\, q_0^2) &=& \DP^{(0)}(q^2,\, q_0^2) +
\!\int_{m^2}^{\infty} \rho^{(\ell)}(\sigma)
\ln\biggl(\frac{\sigma-q^2}{\sigma-q_0^2}
\frac{m^2-q_0^2}{m^2-q^2}\biggr)\frac{d\,\sigma}{\sigma}, \\
\label{Adler_DQCD}
D^{(\ell)}(Q^2) &=& D^{(0)}(Q^2) + \frac{Q^2}{Q^2+m^2}
\int_{m^2}^{\infty} \rho^{(\ell)}(\sigma)
\frac{\sigma-m^2}{\sigma+Q^2} \frac{d\,\sigma}{\sigma}, \\
\label{R_DQCD}
R^{(\ell)}(s) &=& R^{(0)}(s) + \theta(s-m^2) \int_{s}^{\infty}
\rho^{(\ell)}(\sigma) \frac{d\,\sigma}{\sigma}.
\end{eqnarray}
In these equations $\theta(x)$ denotes the unit step--function
($\theta(x)=1$ if $x \ge 0$ and $\theta(x)=0$ otherwise) and
$\rho^{(\ell)}(\sigma)$ is the $\ell$--loop spectral density:
\begin{equation}
\label{RhoGen}
\rho^{(\ell)}(\sigma) = \frac{1}{\pi} \frac{d}{d\,\ln\sigma}\,
\mbox{Im}\lim_{\varepsilon \to 0_{+}} p^{(\ell)}(\sigma-i\varepsilon)
= \frac{1}{\pi}\, \mbox{Im}\lim_{\varepsilon \to 0_{+}}
d^{(\ell)}(-\sigma-i\varepsilon)
= - \frac{d}{d\,\ln\sigma}\, r^{(\ell)}(\sigma),
\end{equation}
with $p^{(\ell)}(q^2)$, $d^{(\ell)}(Q^2)$, and~$r^{(\ell)}(s)$ being the
$\ell$--loop strong corrections to functions~(\ref{PDef}),
(\ref{AdlerDef}), and~(\ref{RDef}), respectively
(see Refs.~\cite{DQCD1, DQCD2} for the details).

It is worthwhile to note that integral representations
(\ref{P_DQCD})--(\ref{R_DQCD}) automatically embody all the
nonperturbative constraints (including the correct analytic properties in
the kinematic variable) that Eqs.~(\ref{DispRelP})--(\ref{DispRelR})
impose on the functions on hand. For~example, dispersion
relation~(\ref{DispRelD}) implies that the Adler function vanishes in the
infrared limit ($D(Q^2) \to 0$ at $Q^2 \to 0$) and possesses the only cut
along the negative semiaxis of real~$Q^2$ starting at the hadronic
production threshold~\mbox{$Q^2 \le -m^2$} (preliminary formulation of the
Dispersive approach to QCD, which accounts for the second constraint only,
was discussed in Ref.~\cite{DQCDPrelim}).

It is worth mentioning also that integral representations
(\ref{P_DQCD})--(\ref{R_DQCD}) were obtained by making use of only the
dispersion relations (\ref{DispRelP})--(\ref{DispRelR}) and the fact that
the strong correction~$d(Q^2)$ vanishes in the ultraviolet
asymptotic~$Q^2\to\infty$. Neither additional approximations nor
model--dependent assumptions were involved in the derivation of
Eqs.~(\ref{P_DQCD})--(\ref{R_DQCD}), see Refs.~\cite{DQCD1, DQCD2} for the
details. It is worthwhile to note that the hadronic vacuum polarization
function~(\ref{P_DQCD}) agrees with relevant lattice simulation data
(e.g., Ref.~\cite{PLat}) and the Adler function~(\ref{Adler_DQCD}) agrees
with corresponding experimental prediction, see Refs.~\cite{DQCD1, DQCD2}
(as well as Ref.~\cite{PRD77}) for the details.

In general, there is no unique way to calculate the spectral
density~(\ref{RhoGen}) (see Refs.~\cite{PRD62, Review}). Nonetheless, the
perturbative contribution to Eq.~(\ref{RhoGen}) can be obtained by making
use of perturbative expressions for the strong
corrections~$p\ind{(\ell)}{pert}(q^2)$, $d\ind{(\ell)}{pert}(Q^2)$,
and~$r\ind{(\ell)}{pert}(s)$ (see, e.g., paper~\cite{CPC} and references
therein):
\begin{equation}
\label{RhoPert}
\rho\ind{(\ell)}{pert}(\sigma) = \frac{1}{\pi} \frac{d}{d\,\ln\sigma}\,
\mbox{Im}\lim_{\varepsilon \to 0_{+}} p\ind{(\ell)}{pert}(\sigma-i\varepsilon)
= \frac{1}{\pi}\, \mbox{Im}\lim_{\varepsilon \to 0_{+}}
d\ind{(\ell)}{pert}(-\sigma-i\varepsilon)
= - \frac{d}{d\,\ln\sigma}\, r\ind{(\ell)}{pert}(\sigma).
\end{equation}

Note that in the massless limit ($m=0$) the integral representations
(\ref{P_DQCD})--(\ref{R_DQCD}) acquire the form
\begin{eqnarray}
\label{P_DQCD0}
&&\quad\quad\DP^{(\ell)}(q^2,\, q_0^2) = -\ln\biggl(\frac{-q^2}{-q_0^2}\biggr) +
\!\int_{0}^{\infty}\! \rho^{(\ell)}(\sigma)
\ln\!\biggl[\frac{1-(\sigma/q^2)}{1-(\sigma/q_0^2)}\biggr]
\frac{d\,\sigma}{\sigma}, \\
\label{AdlerR_DQCD0}
&&D^{(\ell)}(Q^2) = 1 + \int_{0}^{\infty}
\frac{\rho^{(\ell)}(\sigma)}{\sigma+Q^2}\, d\,\sigma, \qquad
R^{(\ell)}(s) = \theta(s) \biggl[1 +  \int_{s}^{\infty}
\!\rho^{(\ell)}(\sigma) \frac{d\,\sigma}{\sigma} \biggr].
\end{eqnarray}
In particular, Eq.~(\ref{P_DQCD0}) expresses the fact that in the massless
limit the hadronic vacuum polarization function $\Pi(q^2)$~(\ref{PDef})
can not be subtracted at the point~$q_0^2=0$. It is worth mentioning also
that for the case of perturbative spectral density
($\rho\ind{(\ell)}{}(\sigma) = \mbox{Im}\; d\ind{(\ell)}{pert}(-\sigma -
i\,0_{+})/\pi$) the massless equations~(\ref{AdlerR_DQCD0}) become
identical to those of the \mbox{so--called} Analytic Perturbation
Theory~\cite{APT} (see also Refs.~\cite{APT1, APT2}). But, as it was
emphasized above, it is essential to keep the hadronic threshold mass~$m$
nonvanishing (see also discussion of this issue in Refs.~\cite{Feynman,
DQCD1, DQCD2}).

In the realistic case ($m \neq 0$) the so--called ``Abrupt kinematic
threshold'' may be employed for the leading--order terms of the
functions~(\ref{RDef})--(\ref{AdlerDef}):
\begin{equation}
\label{PRD_AKT}
\DP^{(0)}(q^2,\, q_0^2) = -\ln\biggl(\frac{m^2-q^2}{m^2-q_0^2}\biggr),
\qquad
D^{(0)}(Q^2) = \frac{Q^2}{Q^2+m^2},
\qquad
R^{(0)}(s) = \theta(s-m^2).
\end{equation}
This equation represents a rather rough approximation, which, nonetheless,
grasps the basic peculiarities of the functions on hand. The
expression~(\ref{PRD_AKT}) was examined in details in Refs.~\cite{DQCD1,
DQCD2, DQCD3} and has been applied to the study of the inclusive
$\tau$~lepton hadronic decay in Refs.~\cite{DQCD2, DQCD3}. The latter has
revealed the significance of the effects due to hadronization. For
example, in the vector channel the leading--order QCD
contribution~(\ref{DeltaQCDDef}) corresponding to Eq.~(\ref{PRD_AKT})
reads $\Delta^{\mbox{\scriptsize (0)}}_{\mbox{\tiny QCD}} = 1 +
\DeltaHad{V}$, where $\DeltaHad{V} \simeq -0.048$, that considerably
exceeds the electroweak correction~$\dpew$ to Eq.~(\ref{RTauTheor}), see
Refs.~\cite{DQCD2, DQCD3} for the details.

More accurate expression for the leading--order terms of the
functions~(\ref{RDef})--(\ref{AdlerDef}) is the so--called
 ``Smooth kinematic threshold'' (e.g., Refs.~\cite{Feynman, QCDAB}):
\begin{eqnarray}
\label{P0SKT}
\DP^{(0)}(q^2,\, 0) \!&=&\! \frac{2}{3} + 2 \biggl(1-\frac{m^2}{q^2}\biggr)\!
\biggl(1 - \frac{\varphi}{\tan\varphi}\biggr), \quad
\sin^2\varphi = \frac{q^2}{m^2}, \\
\label{D0SKT}
D^{(0)}(Q^2) \!&=&\! 1 + \frac{3}{\xi} \biggl\{\! 1 \!+\! \frac{u(\xi)}{2}
\ln\!\Bigl[1 \!+\! 2\xi\Bigl(1 \!-\! u(\xi)\!\Bigr)\!\Bigr]\!\!\biggr\},
\quad\! u(\xi) \!=\! \sqrt{1+\xi^{-1}},
\quad\! \xi=\frac{Q^2}{m^2}, \qquad \quad \\
\label{R0SKT}
R^{(0)}(s) \!&=&\! \theta(s-m^2)\biggl(1-\frac{m^2}{s}\biggr)^{\!\!3/2},
\end{eqnarray}
see also papers~\cite{DQCD3, DQCD4} and references therein. Here the
effects due to hadronization appear to be even more pronounced than in the
aforementioned case, see Refs.~\cite{DQCD3, DQCD4, Prep} for the details.

\section{Inclusive $\tau$~lepton hadronic decay within Dispersive approach}

Let us proceed now to the description of inclusive $\tau$~lepton hadronic
decay within Dispersive approach~\cite{DQCD1, DQCD2}. This analysis
retains the effects due to hadronization (in other words, the
expressions~(\ref{P_DQCD})--(\ref{R_DQCD}) are used instead of their
perturbative approximations and the hadronic threshold mass~$m$ is kept
nonvanishing). The leading--order terms (\ref{P0SKT})--(\ref{R0SKT}) are
also employed.

Eventually, within the approach on hand the quantity
$\DeltaQCD{V/A}$~(\ref{DeltaQCDDef}) acquires the following
form (see Refs.~\cite{DQCD3, DQCD4, Prep} for the details):
\begin{eqnarray}
\label{DeltaQCD_DQCD_ST}
\DeltaQCD{V/A} &=& \sqrt{1-\zva}\,
\Bigl(1 + 6\zva - \frac{5}{8}\zva^{2}
+\frac{3}{16}\zva^{3}\Bigr)
\nonumber \\[-1mm] && \hspace*{-7.5mm}
-3\zva \Bigl(\!\!1 + \frac{1}{8}\zva^{2} - \frac{1}{32}\zva^{3}\! \Bigr)
\ln\biggl[\frac{2}{\zva}\Bigl(\!1\!+\!\sqrt{1-\zva}\Bigr)-1\biggr]\!
+\!\! \int_{m\va^{2}}^{\infty}\!\!\!H\!\Bigl(\frac{\sigma}{M_{\tau}^{2}}\Bigr)\,
\rho(\sigma)\,\frac{d \sigma}{\sigma}\,, \qquad
\end{eqnarray}
where $H(x) = g(x)\,\theta(1-x) + g(1)\,\theta(x-1) - g(\zeta\va)$, $g(x)
= x (2 - 2x^2 + x^3)$, $m_{\mbox{\tiny V}}^{2} \simeq
0.075\,\mbox{GeV}^2$, $m_{\mbox{\tiny A}}^{2} \simeq 0.288\,\mbox{GeV}^2$,
and $\zeta\va = m\va^{2}/\MTau^{2}$. For the spectral
density~$\rho(\sigma)$ the model~\cite{DQCD3, DQCD4}
\begin{equation}
\label{RhoDef}
\rho(\sigma) = \frac{4}{\beta_{0}}\frac{1}{\ln^{2}(\sigma/\Lambda^2)+\pi^2} +
\frac{\Lambda^2}{\sigma}
\end{equation}
(see also papers~\cite{PRD62, Review} and references therein) is used in
this analysis. The first term in the right--hand side of
Eq.~(\ref{RhoDef}) is the one--loop perturbative contribution, whereas the
second term represents intrinsically nonperturbative part of the spectral
density.

\begin{figure}[t]
\includeplots{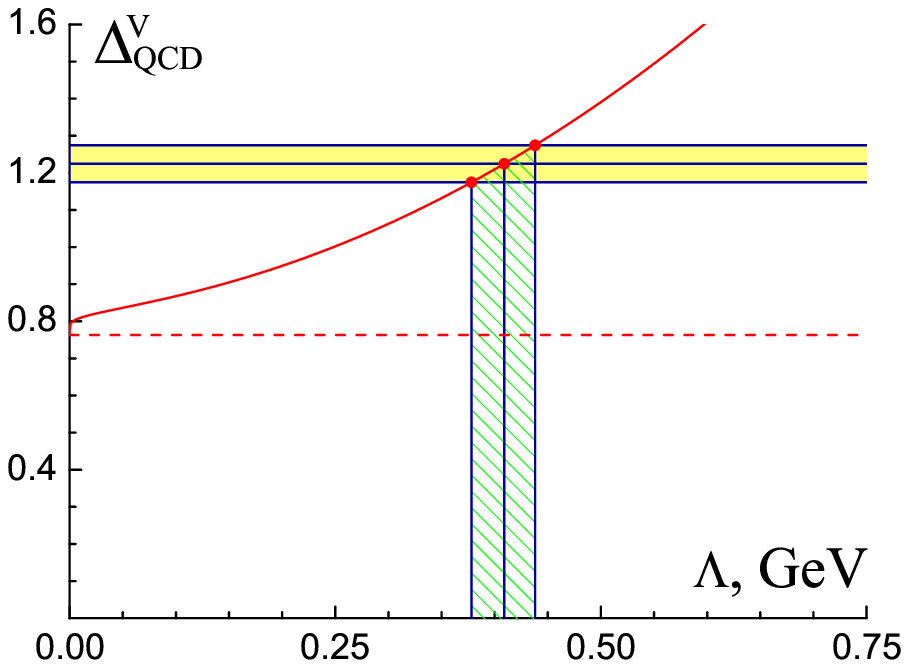}{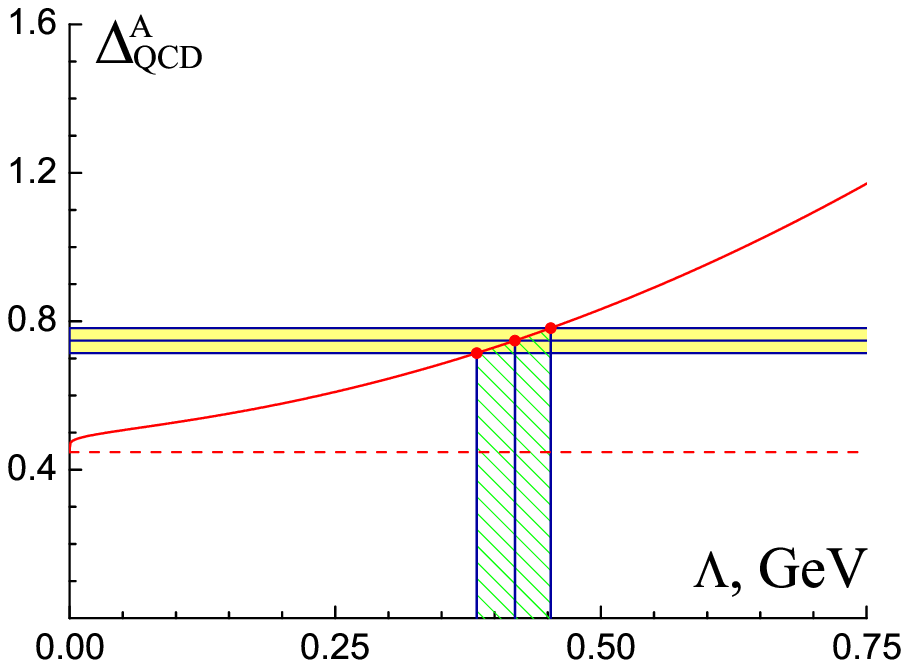}
\caption{Juxtaposition of expression
$\Delta\inds{\!\mbox{V/A}}{QCD}$~(\protect\ref{DeltaQCD_DQCD_ST}) (solid
curves) with relevant experimental data~(\protect\ref{DeltaQCDExp})
(horizontal shaded bands). The solutions for QCD scale parameter~$\Lambda$
are shown by vertical dashed bands.}
\label{Plot:DQCD_ST}
\end{figure}

The comparison of obtained result~(\ref{DeltaQCD_DQCD_ST}) with
experimental data~(\ref{DeltaQCDExp}) gives nearly identical solutions for
the QCD scale parameter~$\Lambda$ in both channels, see
Fig.~\ref{Plot:DQCD_ST}. Namely, $\Lambda = (408 \pm 30)\,$MeV for vector
channel and $\Lambda = (418 \pm 35)\,$MeV for axial--vector one.
Additionally, both these values agree with the aforementioned perturbative
solution for vector channel. It is worth mentioning also that the use of
OPAL data on $\tau$~lepton hadronic decay~\cite{OPAL} yields quite similar
results~\cite{Prep}.

\section{Conclusions}

The theoretical description of inclusive $\tau$~lepton hadronic decay is
performed in the framework of Dispersive approach to QCD. The significance
of effects due to hadronization is convincingly demonstrated. The approach
on hand proves to be capable of describing experimental data on
$\tau$~lepton hadronic decay in vector and axial--vector channels. The
vicinity of values of QCD scale parameter obtained in both channels bears
witness to the self--consistency of developed approach.

\medskip \noindent
The author is grateful to D.$\,$Boito, P.$\,$Colangelo, M.$\,$Davier,
F.$\,$De$\,$Fazio, A.$\,$Francis, and S.$\,$Menke for the stimulating
discussions and useful comments.$\!$ This work is supported by grant
JINR-12-301-01.

\end{document}